%% file: CF4DSP-v3.tex
\documentclass{article}
\usepackage{spconf,algorithm,algpseudocode,subfigure,amsfonts, amsmath, graphicx,url,cite}

\include{def}

\def\ic{\mathrm{IC}_{50}}
\allowdisplaybreaks

\title{From Gene Expression to Drug Response: A Collaborative Filtering Approach}

\name{Cheng Qian$^{\ddag}$ \qquad Nicholas D. Sidiropoulos$^{\ddag}$ \qquad Magda Amiridi$^{\ddag}$ \qquad Amin Emad$^*$
\thanks{Email: alextoqc@gmail.com (C. Qian), nikos@virginia.edu (N. D. Sidiropoulos), ma7bx@virginia.edu (M. Amiridi) and amin.emad@mcgill.ca (A. Emad)}
}

\address{$^{\ddag}$ Department of Electrical and Computer Engineering, University of Virginia, VA, USA \\
$^*$ Department of Electrical and Computer Engineering, McGill University, Montreal, QC, Canada}

\begin{document}

\maketitle

\begin{abstract}
Predicting the response of cancer cells to drugs is an important problem in pharmacogenomics. Recent efforts in generation of large scale datasets profiling gene expression and drug sensitivity in cell lines have provided a unique opportunity to study this problem. However, one major challenge is the small number of samples (cell lines) compared to the number of features (genes) even in these large datasets.
We propose a collaborative filtering (CF) like algorithm for modeling gene-drug relationship to identify patients most likely to benefit from a treatment. Due to the correlation of gene expressions in different cell lines, the gene expression matrix is approximately low-rank, which suggests that drug responses could be estimated from a reduced dimension latent space of the gene expression. Towards this end, we propose a joint low-rank matrix factorization and latent linear regression approach. Experiments with data from the Genomics of Drug Sensitivity in Cancer database are included to show that the proposed method can predict drug-gene associations better than the state-of-the-art methods.
\end{abstract}

\begin{keywords}
    Gene expression, drug response, linear regression, collaborative filtering, Genomics of Drug Sensitivity in Cancer (GDSC).
\end{keywords}

\section{Introduction}
Selecting the right drugs is critical for cancer survival \cite{genedrug2003}, but existing methods that predict a  patient's response to a particular drug are not reliable enough. Resistance to chemotherapy is a major issue, as time is of essence in many cases. Therefore, it is of great interest to construct predictive models of chemotherapy response that physicians can use to prescreen the most promising treatment options. In recent years, the field of pharmacogenomics has emerged as a very promising area with challenging problems that can benefit from more attention from the signal processing community.   

Several large-scale studies have been recently conducted to measure the gene expression (i.e. transcriptomic) profile of hundreds of cell lines and their sensitivity to tens to hundreds of different drugs \cite{ccle,gdsc,moa}. The results of these studies, which are available in databases such as the Cancer Cell Line Encyclopedia (CCLE) \cite{ccle}, the Genomics of Drug Sensitivity in Cancer (GDSC) \cite{gdsc}, and the Cancer Therapeutics Response Portal (CTRP) \cite{moa}, bring predictive models linking gene expression to drug response closer within reach.

Numerous drug sensitivity prediction algorithms have been proposed to characterize the relationship between transcriptomic information and drug response \cite{method0,method1,method2,progeni,44methods,ml}.  
Emad \emph{et al.} recently proposed a gene prioritization method called Prioritization of Genes Enhanced with Network Information (ProGENI) to rank genes that are closely related to a phenotype \cite{progeni}. With the ranked genes, the authors employed a kernel support vector machine (SVM) for drug sensitivity prediction, and showed that ProGENI--identified genes can better predict drug response compared to genes identified by other widely used prioritization methods such as Pearson correlation and Elastic Net (EN).
In \cite{44methods}, through a collaborative effort between the National Cancer Institute (NCI) and the Dialogue on Reverse Engineering Assessment and Methods (DREAM) project, a comparison of 44 different drug response prediction methods was undertaken, among which the Bayesian multitask multiple Kernel learning exhibited the best prediction performance. However, the training was based on just 35 samples, which seems very limited. To handle the cases where the number of genes is greater than that of cell lines, the prevailing methods rely on some sort of sparse regression for gene selection, to help resolve the underdeterminacy that arises in even the simplest linear prediction models \cite{ompdrug,endrug}.

In this paper, we take a different approach. Motivated by the observation that the gene expression matrix is approximately (very) low-rank, instead of relying on gene selection to obtain a well-posed problem, we propose a collaborative filtering (CF) approach based on joint low-rank biased matrix factorization and linear prediction from the latent space. 
It is worth highlighting that unlike existing methods that ignore the bias in the expression of different genes, CF takes this bias into account, which results in a more accurate model.
We provide preliminary results that corroborate the effectiveness of the proposed method using real data from GDSC. 


\section{Proposed Method}
One major challenge, even in large databases such as GDSC \cite{gdsc}, is the large number of features (tens of thousands of genes) compared to the number of samples (hundreds of cell lines).
Therefore, the prediction of drug response from gene expression is inherently under-determined. In the literature, the common way to deal with this problem is to judiciously select a small number of transcriptomic features through sophisticated feature selection methods such as sparse regression or other gene ranking strategies that utilize prior knowledge in the form of protein-protein interactions (PPI's) and genetic interactions \cite{progeni}. 
The existing data sets contain both gene expression data of different cell lines and their response to different drugs, where the response of each drug is only measured for a subset of the cell lines. 
The experimentally measured gene expression data is naturally noisy and is not necessarily `centered'. We propose to model the intrinsically low dimensionality of the gene expression data while taking noise and bias into consideration, using a new method based on collaborative filtering.

One way to tackle biases in the gene expression measurements is to model the gene expression value $g_{ij}$ as
\begin{align}\label{biase2}
g_{ij} = \tilde{g}_{ij} + \beta_j + n_{ij}
\end{align}
where $\tilde{g}_{ij}$ denotes the actual gene expression, $n_{ij}$ is the additive noise which here we assume to be Gaussian distributed with zero mean, and $\beta_j$ is the bias of the $j$th gene.
The matrix form of \eqref{biase2} is given by
\begin{align}\label{biase3}
\G = \tilde{\G} + \1\bbe^T + \N \in\bR^{M\times L}
\end{align}
where $(\cdot)^T$ is the transpose, $\1$ is a vector of  length $M$ with all elements equal to $1$, $M$ is the number of training samples (cell lines), $L$ is the number of transcriptomic features in a cell line, $\G(i,j)=g_{ij}$, $\tilde{\G}(i,j)=\tilde{g}_{ij}$ and $\N(i,j)=n_{ij}$. 

\begin{figure}
	\centering
	\includegraphics[width=1\linewidth]{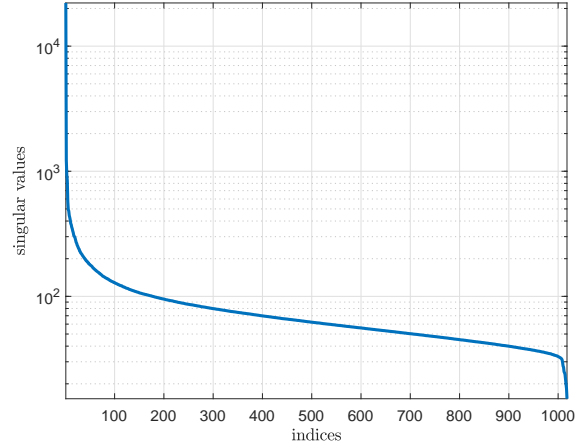}
	\caption{Singular values of the gene expression matrix from GDSC}
	\label{fig:svd}
\end{figure}

To continue, we bring forth our motivation by an example shown in Fig. \ref{fig:svd}, where singular values of a gene expression matrix from the GDSC data set  are plotted. This matrix contains the expression of $17,737$ genes in $1018$ cell lines. Fig. \ref{fig:svd} shows that the gene expression matrix is dominated by a few principal components, indicating that gene expressions of different cell lines are strongly correlated and the gene expression matrix is approximately low-rank. Thus, we have
\begin{align}
\tilde{\G} \approx \A\B^T
\end{align}
where $\A = [\a_1,\cdots,\a_M]^T\in\bR^{M\times F}$ and $\B = [\b_1,\cdots,\b_L]^T$ $\in \bR^{L\times F}$ are low-rank factors with $F\ll \min(M,L)$. Therefore, \eqref{biase3} can be approximated by
\begin{align}
\G \approx \A\B^T + \1\bbe^T + \N.
\end{align}

The above observation implies that it is not necessary to exploit all the transcriptomic features for drug response prediction. On the contrary, the dimension of $\tilde{\G}$ can be significantly reduced by a dimensionality reduction matrix $\B$. As a follow-up, we propose a novel joint dimensionality reduction and drug response prediction strategy, where the drug response is estimated from the latent space of the gene expression matrix---$\A$. 
Mathematically, we try to solve
\begin{align}\label{cf}
\!\!\!\!\min_{\A,\B,\w,\bbe,e}~ &\big\| \G - \A\B^T - \1\bbe^T\big\|_F^2 
	+ \rho\left\| \A\w - \r + e\1 \right\|_2^2
\end{align}
where $\|\cdot\|_2$ is the $\ell_2$-norm, $\|\cdot\|_F$ is the Frobenius norm, $\r$ is the drug response in the training set, $\w$ and $e$ are parameters for fitting the response from the latent space $\A$ such that $\r=\A\w+e\1$.
The first term in \eqref{cf} models the dimensionality reduction and bias cancellation, the second regularization fits the drug response from the latent space of $\G$, and $\rho$ controls the strength of regularization. In \eqref{cf}, fixing any four variables, the problem for the remaining variable is linear least squares (LS).  We therefore employ an alternating least squares (ALS) strategy to solve \eqref{cf}. 
Specifically, at each iteration, the subproblem w.r.t. $\A$ is
\begin{align}\label{prob:Aalpha}
\min_{\A} 
\left\|
\Y_1
- 
\A
\X_1
\right\|_F^2
\end{align}
where 
$
\Y_1 = 
[\G-\1\bbe^T\;\; \sqrt{\rho}(\r-e\1)]
$, $
\X_1 = 
[\B^T\;\; \sqrt{\rho}\w]
$ and the solution is
\begin{align}\label{Aalpha}
\A = \Y_1 \X_1^T(\X_1\X_1^T)^{-1}
\end{align}
with $(\cdot)^{-1}$ being the matrix inverse.

Since $\A\B^T+\1\bbe^T = [\A\; \1][\B\; \bbe]^T$, $\B$ and $\bbe$ can be updated simultaneously. The associated subproblem is 
\begin{align}
\min_{\B,\bbe} 
\left\|
\G^T
- 
\big[
\B\;\; \bbe
\big]
\X_2
\right\|_F^2
\end{align}
where $\X_2 = [\A\;\; \1]^T$
and the minimum is reached at
\begin{align}\label{Bbeta}
[\B\;\; \bbe] = \G^T\X_2^T(\X_2\X_2^T)^{-1}.
\end{align}
Fixing $(\A,\B,\bbe)$, the update for $\w$ is straightforward, i.e.,
\begin{align}\label{w}
\w = (\A^T\A)^{-1}\A^T(\r - e\1).
\end{align}
Finally, we have
\begin{align}\label{e}
e = \frac{1}{M}\sum_{m=1}^{M} (r_m - \a_m^T\w)
\end{align}
where $\a_m^T$ is the $m$th row of $\A$. 
We iteratively update $\{\A,[\B,\bbe],\w,e\}$ until the algorithm converges. Convergence is monotonic in terms of the cost function, by virtue of the conditionally optimal updates of ALS.

So far, we have shown how to estimate the unknown variables in our model. However, it is still unclear how to use $(\A,\B,\bbe,\w)$ to predict drug response for new patients. It is worth noting that $\A$ and $\w$ are not the parameters of interest, instead, $\B$ and $\bbe$ are the ``meat'' and ``bread''. To explain this point, let us first showcase how to use $\B, \bbe$ for dimensionality reduction of a new cell line $\g \in \bR^{L \times 1}$ (note that we now switch to a using a column vector for the cell line). 
We then solve
\begin{align}
\min_{\a} \| \g - \bbe - \B\a \|_2^2
\end{align}
resulting in the reduced dimension gene expression vector
\begin{align}\label{g}
\hat{\a} = (\B^T\B)^{-1}\B^T(\g - \bbe) \in\bR^{F}.
\end{align}

Comparing \eqref{g} and \eqref{Aalpha}, we see that $\A$ in \eqref{Aalpha} is updated differently since it is a regularized LS that involves $\sqrt{\rho}\r$ and $\sqrt{\rho}\w$ while $\hat\a$ does not.
Nevertheless, for real-data applications, the new gene expression may not rigorously follow the proposed model, which means that $\w$ estimated from ALS is not properly paired with the new cell line $\hat{\a}$. Thus, estimating the response from $\hat\a\w$ is not the best option.

To handle this issue, we need to recalculate $\w$ by using a refined $\A$ obtained in the same manner as \eqref{g}, i.e.,
\begin{align}\label{hatA}
\hat{\A} = (\G - \1\bbe^T)\B(\B^T\B)^{-1}.
\end{align}
Hence, by minimizing $\|\hat{\A}\hat{\w} - \r\|$, we have
\begin{align}\label{hatw}
\hat{\w} = (\hat{\A}^T\hat{\A})^{-1}\hat{\A}^T(\r - e\1).
\end{align}
The drug response is then estimated through
\begin{align}\label{r}
\hat{r} = \hat{\a}^T\hat{\w} + e.
\end{align}

\noindent The overall procedure is summarized in Algorithm \ref{algorithm1}.

\noindent\textit{Remark}:  As we can see from \eqref{Bbeta}, $\bbe$ is not simply a gene expression bias vector; it actually plays an important role in finding an accurate dimensionality reduction matrix $\B$ and assisting drug response estimation from the latent space.

\begin{algorithm}[t]
	\caption{Collaborative Filtering }
	\begin{algorithmic}[1]
		\Function{CF}{$\G,\r,F,\rho$}
		\State Randomly initialize $\B$ and $\w$
		\State Set $\ell=1$
		\While{stopping criterion has not been reached}
		\State $\A \leftarrow \eqref{Aalpha}$
		\State $[\B\;\;\bbe] \leftarrow \eqref{Bbeta}$
		\State $\w \leftarrow \eqref{w}$
		\State $e \leftarrow \eqref{e}$
		\State $\ell = \ell + 1$
		\EndWhile
		\State Refine $\A$ with \eqref{hatA} and $\w$ with \eqref{hatw}
		\State Given a new cell line $\g$, reduce its dimension via \eqref{g} and then compute the drug response using \eqref{r}
		\EndFunction
	\end{algorithmic}\label{algorithm1}
\end{algorithm}

\section{Results}


%


We compare the performance of of our CF-inspired approach with state-of-the-art algorithms using real data from GDSC, which is fully accessible at \url{https://www.cancerrxgene.org/downloads}.
We use the RMA-normalised basal gene expression profiles of the cell lines released on March 2, 2017.
Tthe drug response data that we use was released on March 27, 2017, containing the biochemical half maximal inhibitory concentration ($\ic$) values of different drugs for each cell line.

For most of the cell lines in GDSC, the expression values of approximately $18,000$ genes are available, but the drug that has been measured using the largest number of samples includes approximately $1,000$ samples -- which poses a challenge for training an accurate prediction model. Therefore, it is necessary to prudently select a smaller subset of informative features for training while excluding the irrelevant ones. Toward this end, we first employ the ProGENI algorithm \cite{progeni} to rank the genes for each drug and select the top $500$ genes to construct a smaller-size gene expression matrix. Then we choose $70\%$ of the data samples of a tested drug for training, $10\%$ for validation and $20\%$ for testing.

\begin{table}[t]                                 
	\caption{RMSE Comparison over 10 drugs}\vspace{0.2em}               
	\centering                                               
	\begin{tabular}{c|cccc}                             
		\hline                                                   
		drug name & CF & OMP & IHT & EN \\                       
		\hline                                                   
		SN-38 & \textbf{0.2928} & 0.3173 & 0.3824 & 0.4998 \\             
		\hline                                                   
		TAK-715 & \textbf{0.2286} & 0.2372 & 0.2439 & 0.4604 \\           
		\hline                                                   
		Ruxolitinib & \textbf{0.1875} & 0.2006 & 0.1952 & 0.3605 \\       
		\hline                                                   
		Ispinesib Mesylate & \textbf{0.6293} & 0.6518 & 0.6660 & 1.2147 \\
		\hline                                                   
		BX-912 & \textbf{0.4475} & 0.4467 & 0.4663 & 0.8905 \\            
		\hline                                                   
		Avagacestat & \textbf{0.1228} & 0.1290 & 0.1319 & 0.2571 \\       
		\hline                                                   
		XMD14-99 & \textbf{0.1807} & 0.1866 & 0.1945 & 0.3375 \\          
		\hline                                                   
		PHA-793887 & \textbf{0.5318} & 0.5593 & 0.5509 & 1.0625 \\        
		\hline                                                   
		XMD15-27 & \textbf{0.1648} & 0.1729 & 0.1784 & 0.3242 \\          
		\hline                                                   
		Quizartinib & \textbf{0.3765} & 0.4042 & 0.4123 & 0.7231 \\          
		\hline                                                   
	\end{tabular}                                         
	\label{table:MyTableLabel}                               
\end{table}  

In the first example, we compare CF with orthogonal matching pursuit (OMP), iterative hard-thresholding (IHT) and Elastic Net (EN) in terms of {\underline{relative}} root mean square error (RMSE). We choose 10 drugs (i.e., SN-38, TAK-715, Ruxolitinib, Ispinesib Mesylate, BX-912, Avagacestat, XMD14-99, PHA-793887, XMD15-27, Quizartinib) to examine the performance of different competitors and report their RMSEs. Here, RMSE is averaged through 10 random permutations and is calculated as
$$
\mathrm{RMSE} = \frac{1}{10}\sum_{i=1}^{10} \|\r - \hat\r_i\|_2 / \|\r\|_2
$$
where $\hat\r_i$ contains the drug response estimates of the testing cell lines from the $i$th permutation. Since the rank $F$ and hyper-parameter $\rho$ for CF are unknown, we vary $F$ from 5 to 30 and $\rho$ from 100 to 400, and choose $\{F,\rho\}$ that minimizes the Euclidean distance between the estimated and true drug response vector of the validation set. Simulation results are shown in Table 1, where CF has the smallest RMSE for all 10 drugs. Overall, OMP performs slightly better than IHT while EN has the worst performance.

In the second example, we examine the performance of CF on predicting drug sensitivity. We compare our method with OMP and three classifiers from MATLAB Statistics and Machine Learning Toolbox as baselines, i.e., logistic regression, kernel SVM (K-SVM) with Gaussian kernel function and linear SVM (L-SVM). Here, we do not include IHT and EN because their performance is worse than OMP.
The drug selected for comparison is SN-38. This drug has the largest number of samples in the GDSC data set. There are 989 cell lines tested with this drug, but only 956 of them have associated transcriptomic data, which means that the total number of available samples is 965. 
Note that we define a threshold such that a cell line with $\ic$ value smaller than the threshold is identified as sensitive to the drug; otherwise, resistant. Thus, given a threshold, the data set can be divided into two parts corresponding to drug sensitive and resistant, respectively. 
The $\ic$ values for this drug range from $-8.1319$ to $1.4428$, where the smaller the $\ic$ value, the more sensitive the cell line to this drug. Therefore, we vary the threshold from $-5.5$ to $-2.5$ to compare the prediction performance of different algorithms. Then given a threshold, we choose $70\%$ of the data samples from each of the two parts for training, $10\%$ for validation and $20\%$ for testing, such that the percentage of either resistant or sensitive samples is fixed in training, validation and testing sets. 

Similar to the previous example, we run 10 Monte Carlo simulations with randomly partitioned training/validation/testing sets and report the average prediction accuracy of the testing set defined as
$$
\mathrm{prediction~accuracy} = \frac{\sum_{i=1}^{N}\delta(\ell_i, \hat\ell_i)}{N}
$$
where $\ell_i$ is the drug response label, $\hat\ell_i$ is the estimated label, $N$ is the size of the testing data and $\delta(a,b)=1$ if $a=b$ and $0$ otherwise.
Also notably, for logistic regression and SVM, in consideration of the limited training samples, we employ OMP to further reduce the number of features by solving $\min_{\|\s\|_0\leq 20} \|\G\s - \r\|_2^2$, where the indices of the nonzero elements in $\s$ denote the selected features. 

\begin{figure}
	\centering
	\includegraphics[width=1\linewidth]{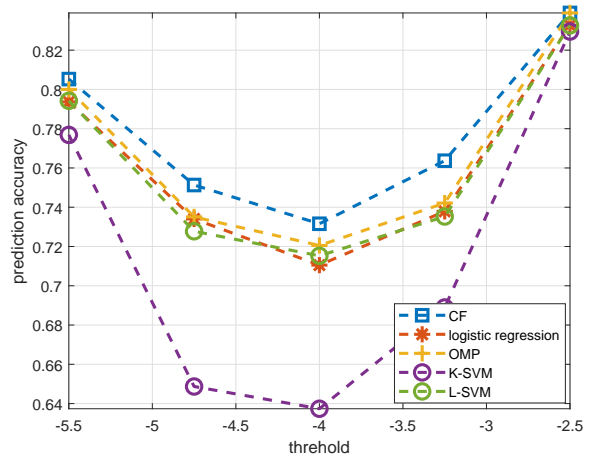}
	\caption{Prediction accuracy comparison on drug sensitivity.}
	\label{fig:predaccu}
\end{figure}

Fig. \ref{fig:predaccu} shows the results, from which we see that CF has the highest prediction accuracy under different thresholds. Its performance is followed by OMP, logistic regression and L-SVM. However, the K-SVM does not work well.

\section{Conclusion}
A novel CF algorithm has been proposed for drug response prediction from gene expression. Simulations validated that CF works better than many sparse regression methods (e.g., OMP, IHT and EN) and classical linear and nonlinear classification algorithms (e.g., logistic regression and SVM). 

The CF method estimates the $\ic$ values rather than the labels of drug sensitive/resistant. This can be valuable for imputing missing $\ic$'s of cell lines for which the drug response is not measured, and for understanding the relationship between gene expression and drug response.  Another important advantage of our CF-based method is that it can be used in the case of incomplete gene expression measurements -- i.e., even when the matrix ${\bf G}$ has many missing entries. The only change in this case is that one needs to use weighted LS or stochastic gradient updates within the main ALS algorithm, as is well-known in the collaborative filtering literature. 

\end{document}

%% file: def.tex
\def\bbe{\boldsymbol{\beta}}

\def\({\left(}
\def\){\right)}
\def\[{\left[\,}
\def\]{\,\right]}

\def\0{\boldsymbol{0}}
\def\1{\boldsymbol{1}}
\def\a{\mathbf{a}}
\def\A{\mathbf{A}}

\def\b{\mathbf{b}}
\def\B{\mathbf{B}}

\def\g{\mathbf{g}}
\def\G{\mathbf{G}}

\def\N{\mathbf{N}}

\def\r{\mathbf{r}}

\def\bR{\mathbb{R}}
\def\s{\mathbf{s}}

\def\w{\mathbf{w}}

\def\X{\mathbf{X}}

\def\Y{\mathbf{Y}}